\documentclass[sigconf]{acmart}

\usepackage{url}
\usepackage{multirow}
\usepackage{array}
\usepackage{amsmath}
\usepackage{amssymb}
\usepackage{booktabs}
\usepackage{graphicx}
\usepackage{flushend}
\usepackage{enumitem}
\usepackage{xcolor}

\definecolor{pink}{rgb}{0.9,0,0.9}
\definecolor{gray}{rgb}{0.95,0.95,0.85}
\definecolor{mauve}{rgb}{0.1,0.7,0.2}
\definecolor{dkgreen}{rgb}{0,0.6,0}

\setcopyright{acmlicensed}

\urldef\dataseturl\url{https://goo.gl/2NMjcF}

\begin{document}

\copyrightyear{2018}
\acmYear{2018}
\setcopyright{acmlicensed}
\acmConference[ICSE'18]{ICSE'18: 40th International Conference on Software Engineering }{May 27-June 3, 2018}{Gothenburg, Sweden}
\acmPrice{15.00}
\acmDOI{10.1145/3180155.3180207}
\acmISBN{978-1-4503-5638-1/18/05}

\title[Traceability in the Wild: Augmenting Incomplete Trace Links]{Traceability in the Wild: \\Automatically Augmenting Incomplete Trace Links}



\author{Michael Rath$^1$, Jacob Rendall$^2$, Jin L.C. Guo$^3$, Jane Cleland-Huang$^2$, Patrick M{\"a}der$^1$}
\affiliation{%
 \institution{Technical University Ilmenau, Ilmenau, Germany$^1$}
 \institution{University of Notre Dame, South Bend, USA$^2$}
 \institution{McGill University, Montreal, Canada$^3$}
}
\email{{michael.rath,patrick.maeder}@tu-ilmenau.de;{jrendal1,JaneClelandHuang}@nd.edu; jguo@cs.mcgill.ca}




\renewcommand{\shortauthors}{M. Rath et al.}

\begin{abstract}
Software and systems traceability is widely accepted as an essential element for supporting many software development tasks. Today's version control systems provide inbuilt features that allow developers to tag each commit with one or more issue ID, thereby providing the building blocks from which project-wide traceability can be established between feature requests, bug fixes, commits, source code, and specific developers. However, our analysis of six open source projects showed that on average only 60\% of the commits were linked to specific issues. Without these fundamental links the entire set of project-wide links will be incomplete, and therefore not trustworthy. In this paper we address the fundamental problem of missing links between commits and issues. Our approach leverages a combination of process and text-related features characterizing issues and code changes to train a classifier to identify missing issue tags in commit messages, thereby generating the missing links. We conducted a series of experiments to evaluate our approach against six open source projects and showed that it was able to effectively recommend links for tagging issues at an average of 96\% recall and 33\% precision. In a related task for augmenting a set of existing trace links, the classifier returned precision at levels greater than 89\% in all projects and recall of 50\%. 
\end{abstract}

%
%

\maketitle

\section{Introduction}\label{sec:introduction}
Traceability provides support for many different software engineering activities including safety analysis, change impact analysis, test regression selection, and coverage analysis \cite{SASTBook, DBLP:conf/re/GotelF94, DBLP:journals/tse/RameshJ01,DBLP:journals/sosym/MaderC13,DBLP:conf/icse/Cleland-HuangGHMZ14,DBLP:journals/software/MaderC15}. Its importance has long been recognized in safety-critical domains, where it is often a prescribed part of the development process \cite{DBLP:conf/refsq/Cleland-HuangHHLM12,DBLP:journals/software/MaderJZC13,DBLP:conf/icse/RempelICSE14,DBLP:conf/eurospi/ReganBMMF14,DBLP:conf/se/RempelM16}. While traceability is relevant to all software development environments \cite{DBLP:journals/tse/RameshJ01,DBLP:journals/software/MaderJZC13,DBLP:journals/ese/MaderE15,DBLP:journals/ese/MaderOM17,DBLP:journals/ese/StahlHB17,DBLP:journals/tse/RempelM17}, the effort needed to manually establish and maintain trace links in non-regulated domains has often been perceived as prohibitively high.


However, with the ubiquitous adoption of version control systems such as \emph{Git}~\cite{git} and \emph{GitHub}, and issue tracking systems such as \emph{Bugzilla} or \emph{Jira}~\cite{jira}, it has become common practice for developers to tag commits with issue IDs. In large projects, such as the ones from the Apache Foundation, this procedure is reflected in the guidelines which state that \emph{``You need to make sure that the commit message contains at least [\ldots] a reference to the Bugzilla or JIRA issue [\ldots]''} \cite{apache_guideline}. Creating such tags establishes explicit links between commits and issues, such as feature requests and bug reports. However, the process is not perfect, as developers may forget, or otherwise fail, to create tags when they make a commit \cite{DBLP:conf/wikis/RomoCH14, Bachmann:2009:SPD}.
While the practice of tagging commits has become popular in open source projects, it is conceptually applicable in any project where version control systems and issue trackers are used. 

In this paper we propose a solution for identifying tags that are missing between commits and issues and augmenting the traceability data with these previously missing links. As shown later in the paper, our observations across six OSS showed that an average of only about 60\% of commits were linked to specific issues. The majority of papers addressing traceability in OSS have focused on directly establishing a complete set of links between issues and source code.  In contrast, we focus on generating the missing links at the commit level.  This has the primary advantage of providing traceability support within the natural context in which developers are creating trace links. Our approach leverages existing tags, as well as information related to the commit process itself and also textual similarities between commit messages, issue descriptions, and code changes. We use these attributes to train a classifier
to identify tags that are missing from commit messages.  Furthermore, we set a critical constraint on our work that the classifier must be populated, trained, and then utilized with a simple ``button press'' in order to make it practical in an industrial setting. 

Low level links between commits and issues provide the building blocks for inferring project-wide traceability between improvements, bug reports, source code, test cases, and commits, and also allow associations to be established between the issues and developers \cite{DBLP:conf/refsq/SeilerP17}. Augmenting the set of trace links between commits and issues, therefore results in a more complete set of project-wide trace links. This enables more accurate support for tasks such as defect prevention \cite{DBLP:journals/tse/RempelM17}, change impact analysis, coverage analysis, and even provides enhanced support for building recommendation systems to identify appropriate developers for fixing bugs \cite{DBLP:conf/icse/AnvikHM06}.

We train and evaluate our approach on six open-source projects in order to address three key research questions:\par
\noindent{\bf RQ1: } Is the link classifier able to accurately reconstruct issue tags during the commit process?\par
\noindent{\bf RQ2: } Is the link classifier able to precisely augment an existing set of incomplete commit to issue links in a fully automated way?\par
\noindent{\bf RQ3: } Is the link classifier able to recommend additional tags?\par

The remainder of the paper is structured as follows. Section \ref{sec:Fundamentals} introduces the artifacts, case projects, process model, and stakeholder model, that form the fundamentals of our approach. Section \ref{sec:LinkClassifier} describes the elements of our classifier. Section \ref{sec:Data} describes the six projects in our study. Sections \ref{sec:KnownLinks} and \ref{sec:UnknownLinks} describe scenarios and experiments associated with recommending tags for commits, augmenting existing sets of trace links, and constructing trace links for commits with no tags.  Finally sections \ref{sec:Related} to \ref{sec:Conclusion} discuss related work, threats to validity, and conclusions.

\section{Fundamentals}
\label{sec:Fundamentals}
We first introduce a motivating example and describe the artifacts, project environments, and the process and stakeholder models that form the fundamentals of our approach.

\subsection{Motivating Example}
Figure~\ref{fig:example_improvement} depicts the improvement request, \texttt{GROOVY-5223},\footnote{\url{https://issues.apache.org/jira/browse/GROOVY-5223}} retrieved from the \textsc{Groovy} project's issue tracker, JIRA \cite{jira}.
The request consists of a unique issue ID, a short summary, a longer textual description, time stamps for issue creation and resolution, the issue's current status, and information about its resolution. This particular improvement requests an enhancement to an existing feature concerning class loading at byte code level.
Figure~\ref{fig:example_bug} shows a bug report \texttt{GROOVY-5082}\footnote{\url{https://issues.apache.org/jira/browse/GROOVY-5082}} for the same project. It includes the same fields as the improvement, except that the type is specified as a bug. In this case, the bug describes a problem with byte code generation for the groovy language. 
Finally, Figure~\ref{fig:example_commit} shows an example of a  commit\footnote{\url{http://goo.gl/pBy6Nw}} submitted to the Git \cite{git} version control system. A commit (change set) includes a unique commit hash value, a message describing its purpose, the time stamp when it was submitted, and finally a list of files modified by the change set.
\begin{figure}
  \small
  \centering
  \fbox{\parbox{0.97\columnwidth}{
    \makebox[1.8cm][l]{\textbf{Issue ID:}} \texttt{GROOVY-5223}\hfill \makebox[1.8cm][l]{\textbf{Type:}} \makebox[2.2cm][l]{Improvement}\\
    \makebox[1.8cm][l]{\textbf{Summary:}} [GROOVY-5223] Bytecode optimizations: make use of LDC for class literals\\
    \makebox[1.8cm][l]{\textbf{Description:}} Class literals are currently loaded using generated \$get\$class\$ methods which increase bytecode size and may prevent some optimizations. In most situations though, we may use the LCD bytecode instruction to load the class literal.\\
    \makebox[1.8cm][l]{\textbf{Status:}} Closed\hfill\makebox[1.7cm][l]{\textbf{Created:}} \makebox[2.2cm][l]{30/Dec/11 09:59}\\
    \makebox[1.8cm][l]{\textbf{Resolution:}} Fixed\hfill\makebox[1.8cm][l]{\textbf{Resolved:}} \makebox[2.2cm][l]{03/Jan/12 02:29}
    }}
    \vspace{-6pt}
  \caption{Example of an improvement in Jira issue tracker}
  \label{fig:example_improvement}
  \vspace{-6pt}
\end{figure}
\begin{figure}
  \small
  \centering
  \fbox{\parbox{0.97\columnwidth}{
    \makebox[1.8cm][l]{\textbf{Issue ID:}} \texttt{GROOVY-5082}\hfill \makebox[1.8cm][l]{\textbf{Type:}} \makebox[2.2cm][l]{Bug}\\
    \makebox[1.8cm][l]{\textbf{Summary:}} [GROOVY-5082] Sometimes invalid inner class reference left in .class files produced for interfaces\\
    \makebox[1.8cm][l]{\textbf{Description:}} Compile this: \emph{[...]} Upon javap'ing the result we see this InnerClass attribute: \emph{...} to X\$1. But there is no X\$1 produced on disk.\emph{...}\\
    \makebox[1.8cm][l]{\textbf{Status:}} Closed\hfill\makebox[1.7cm][l]{\textbf{Created:}} \makebox[2.2cm][l]{17/Oct/11 13:45}\\
    \makebox[1.8cm][l]{\textbf{Resolution:}} Fixed\hfill\makebox[1.8cm][l]{\textbf{Resolved:}} \makebox[2.2cm][l]{01/Feb/12 11:10}
    }}
    \vspace{-6pt}
  \caption{Example of a resolved bug in Jira issue tracker}
  \label{fig:example_bug}
  \vspace{-6pt}
\end{figure}
\begin{figure}
\small
\centering
\fbox{\parbox{0.97\columnwidth}{
\setlength{\fboxrule}{1pt} 
\makebox[1.6cm][l]{\textbf{Hash:}}\texttt{b1bb2abfde414950238ff4d895bf5e182793500a}\hfill\\
\makebox[1.6cm][l]{\textbf{Message:}}GR{\color{blue!90!black!60}\framebox[8pt]{\normalcolor O}}VY-5082: remove synthetic interface loading helper class in case it is not used\hfill \\
\makebox[1.6cm][l]{\textbf{Committed:}}Feb 1, 2012\hfill\\
\makebox[1.6cm][l]{\textbf{Files:}}\texttt{\scriptsize src/main/org/codehaus/groovy/classgen/AsmClassGenerator.java}
}}
\vspace{-6pt}
\caption{Example for a commit in Git}
\label{fig:example_commit}
\vspace{-12pt}
\end{figure}


The common way to establish a trace link between a commit and an issue is by placing the unique issue id, i.e., \texttt{GROOVY-5082} in this example, into the beginning of the commit message. 
However, a close examination of the commit message in this example shows that the committer made a subtle mistake and misspelled the issue key for the bug that was being fixed (omitting an \textbf{O}). As a result, traditional trace link construction techniques that rely upon matching the key to an issue will fail to create a trace link.

However, even without a valid issue key, there are numerous clues to suggest that the commit should be associated with the reported bug. First, the bug description exhibits textual similarity to the commit message as well as to the text in the changed file \texttt{AsmClassGenerator.java}. Second, the commit was submitted on the same date that the issue was resolved, and finally,  the person (obfuscated for privacy reasons) who submitted the commit was also responsible (i.\,e. the assignee) for resolving the issue. Taken together, these observations provide some degree of evidence that the commit and bug should be linked. This example illuminates the thinking behind our proposed solution.  We build a classifier that leverages all of this information, plus additional attributes, to learn which issues should be tagged to each specific commit.

\subsection{Software Artifacts and their Relations}
While version control systems and issue trackers have several types of artifacts, our approach leverages three of them to construct missing commit links. These are \emph{issues}, \emph{commits} (i.e., change sets), and \emph{source code files}.

\textbf{Issues:} Our model uses issues collected from the Jira issue tracking system.  While there are several types of \emph{issues}, we focus on \emph{improvements} and \emph{bugs} which are the most commonly occurring ones.  An improvement represents an enhancement to an existing feature in the software, while a bug describes a problem, which impairs or prevents its correct functionality. In the remainder of the paper, the term \emph{issue} is used in reference to both improvements and bugs. 
Independent of their actual type, all issues share the following properties: a unique \emph{issue ID}, a \emph{summary} providing a brief one-line synopsis of the issue, and a more extensive explanation provided in the \emph{description}. Further, every issue has a temporal life cycle -- it is \emph{created} at a given point in time and later \emph{resolved}, and may be assigned to an author, responsible for its resolution.

\textbf{Commit (Change Set):} In Git version control, changes are organized as atomic \emph{commits}. A commit bundles together all modified files and is uniquely identified by a \emph{hash} value. It includes properties concerning the person who made the change, a time stamp (committed date), and a commit message stating the purpose of the change.

\textbf{Source Code:} Multiple types of files are associated with a software project including \emph{source code files}, \emph{documentation}, \emph{examples}, and \emph{tests}. In this paper we focus on source code files which are explicitly linked to commit messages.  The source code files provide support for our primary goal of establishing links between commit messages and issues.

\begin{figure}
  \includegraphics[trim={0 0.9cm 0 1.2cm}, clip]{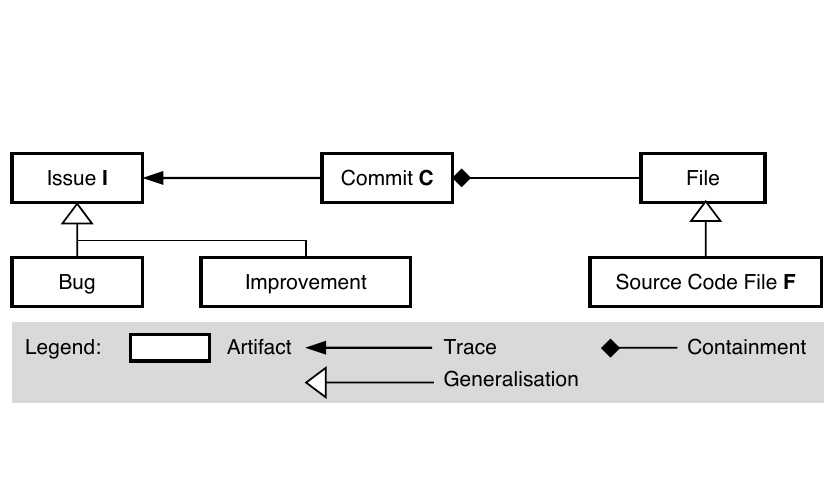}
  \caption{Studied artifact model with issues $\mathcal{I}$, commits $\mathcal{C}$, source code files $\mathcal{F}$, and their relations.}
  \label{fig:artifacts}
  \vspace{-12pt}
\end{figure}

\textbf{Relations:} We are also interested in relations between the three fundamental types of artifacts in our model. A commit atomically bundles one or more source files together. This containment relation between commit and source code artifacts is a natural result of submitting the change set to the version control system.
Further, as previously explained, trace links are explicitly created between issues and commits when a developer tags a commit with a valid issue ID. We denote an issue as \emph{linked}, if there is at least one trace link from the issue to a commit.  Issues without any links are termed \emph{non linked}. Figure~\ref{fig:artifacts} depicts the three artifacts as well as their structural interactions.
We denote $\mathcal{I} = \mathcal{I}_{Bug} \cup \mathcal{I}_{Imp}$ as the set of issues (bugs and improvements), $\mathcal{C}$ the set of commits, and $\mathcal{F}$ the set of source code files in a project. The function $is\_linked: \mathcal{C} \times \mathcal{I} \rightarrow \{0, 1\}$ returns $1$ if an explicit link exists, and $0$ otherwise. The function $mod(C): C \rightarrow  F_{C}$ with $F_{C} \subseteq \mathcal{F}, C \in \mathcal{C}, $ calculates this set for a given commit. A source code file may be part of multiple commits.


\subsection{Studied Projects}
For our study, we selected six projects from diverse domains, that utilized both Git and Jira. They included: build automation (\textsc{Maven} (Ma)), databases (\textsc{Derby} (De), \textsc{Infinispan} (In)), languages (\textsc{Groovy} (Gr), \textsc{Pig} (Pi)), and a rule engine (\textsc{Drools} (Do)), primarily selected because each of these projects has existed for several years, has a non-trivial number of commits and issues, and largely followed the practice of tagging commits with issue IDs.  We analyzed each of the projects to gain an understanding of the numbers of links that existed between commit messages and issues. Further, we analyzed the number of issues that were linked to exactly one commit (1:1), two or more disjoint commits (1:N), or had no links. Results are reported in Table \ref{tab:issue_to_commit_linkage}. For example, of the 2,638 bug-related issues in the \textsc{Derby} project, 1,093 were linked to only one commit, 273 were linked to multiple commits, and 1,272 had no associated commits.  Across all of the projects approximately 43.3\% of improvements and 42.4\% of bugs have no commits associated with them. 

\begin{table}
  \renewcommand{\arraystretch}{0.8}
  \caption{Existing bug and improvement to commit link characteristics in the studied project.}
  \label{tab:issue_to_commit_linkage}
  \centering
  \begin{tabular}{@{}ll rrrrrrr@{}}
    \toprule
                         &           & \multicolumn{6}{c}{Project} \\
    \cmidrule{3-8}
    Link Type                 & Profile   & De & Dr  & Gr  & In    & Ma & Pi   \\
    \midrule
    \multirow{2}{*}{1:1} & Bug       & 1093  &  1040   &  1609   & 1714 & 609   & 1066  \\
                         & Imp.      &  399  &   61    &   459   & 393  & 222   & 313   \\
    \midrule
    \multirow{2}{*}{1:n} & Bug       &  273  &   236   &  422    & 254  & 99    & 87    \\
                         & Imp.      &  225  &    28   &   179   & 96   & 46    & 48    \\
    \midrule
    \multirow{2}{*}{Non} & Bug       & 1272  &   605   &  1667   & 994  & 769   & 944   \\
                         & Imp.      &  730  &   42    &  392    & 157  & 313   & 251   \\ 
    \bottomrule
\end{tabular}
\end{table}

Table~\ref{tab:commit_to_issue_linkage} depicts a similar analysis from the perspective of the commits.  It reports the number of commits with links to issues for the selected projects. Again, we analyzed the distribution of 1:1 links, 1:N links and non linked commits.  In the \textsc{Derby} project, of the 3,735 commits, 1,657 linked to only one bug, 1,350 to only one improvement, 175 linked to multiple bugs or improvements, and 553 commits had no links. However, across all of the projects approximately 48\% of the commits were not linked to any issue.  Furthermore, there was significant variance across the six projects with only 15\% of commits in \textsc{Derby} having no links compared to approximately 76\% of unlinked commits in \textsc{Maven}. Clearly, different practices exist across different projects, leading to huge disparities in the extent to which issue tags are added to commit messages.
\begin{table}
  \renewcommand{\arraystretch}{0.8}
  \caption{Existing commit to bug and improvement link characteristics in the studied project.}
  \label{tab:commit_to_issue_linkage}
  \begin{tabular}{@{}ll rrrrrr@{}}
    \toprule
    &         &       \multicolumn{6}{c}{Project}       \\
    \cmidrule{3-8}
    Link Type                   & Profile & De   &   Dr &    Gr &    In &    Ma &    Pi \\ \midrule
    \multirow{2}{*}{1:1}    & Bug     & 1657 & 1559 & 2423 & 2223 &  781 & 1206 \\
    & Imp.    & 1350 &  195 &  903 &  671 &  335 &  437 \\ \midrule
    1:n                    & both    & 175  &  138 &  139 &   95 &   48 &   38 \\ \midrule
    Non                  &         & 553  & 2947 & 4740 & 1479 & 3614 &   64 \\
    \bottomrule
  \end{tabular}
\end{table}

One of the primary goals of our work, is to establish at least one link for each commit. As the majority of commits link to a single issue, we only attempt to generate links for currently unlinked commits.  For commits without links there are two viable cases -- first that an appropriate issue exists and a link can be generated, and second that no appropriate issue exists for the commit. 

\subsection{Process Model}
\begin{figure*}
  \includegraphics[trim={0cm 10.2cm 0cm 1cm},clip]{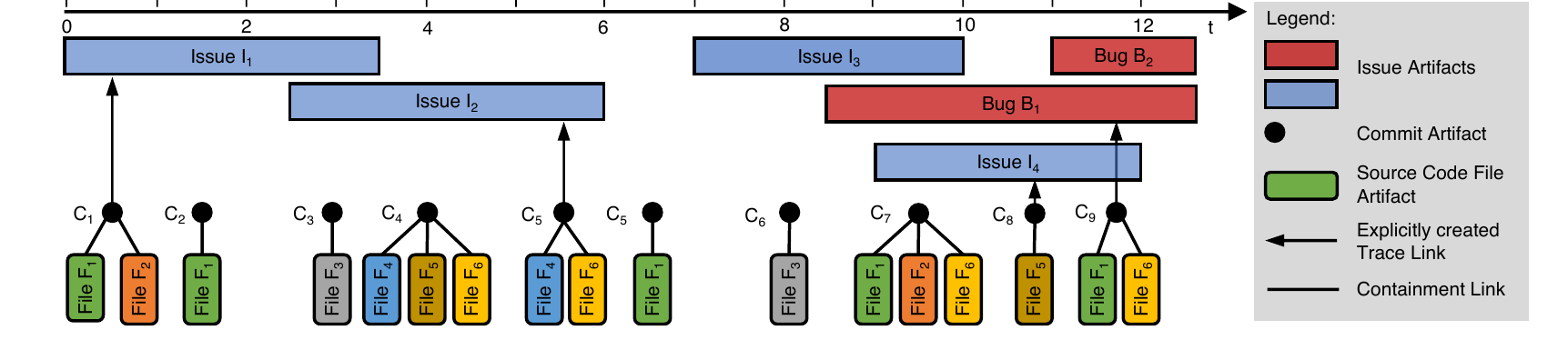}
   \caption{Temporal and structural relations between issue ($\mathcal{I}$), commit ($\mathcal{C}$), and source code file ($\mathcal{F}$) artifacts.}
  \label{fig:timeline}
\end{figure*}

As previously explained, our approach leverages clues from the development process to aid in the generation of links. First we observe that the software development process is time dependent: bugs and improvements are constantly created and resolved, and commits are submitted to the version control system. Figure~\ref{fig:timeline} exemplifies this scenario.
It contains six issues $\mathcal{I} = \{I_1 \ldots I_4, B_1, B_2\}$, nine commits $\mathcal{C} = \{C_1 \ldots C_9\}$, and six source code file artifacts $\mathcal{F} = \{F_1 \ldots F_6\}$. The issue artifacts and commits are ordered across a time line. In this example, the issues $I_1, I_2, I_4, B_1$, as well as commits $C_1, C_5, C_8$, and $C_9$ are linked, e.\,g. $is\_linked(C_1, I_1) = 1$. The figure also shows the relation between issues and commits according to the time line. We define the functions $created: I \rightarrow \mathbb{N}$ and $resolved: I \rightarrow \mathbb{N}, I \in \mathcal{I}$, which returns the point in time when the issue was created and respectively resolved. During this time, the issue is considered to be unfinished and source code modifications are required in order to implement the improvement or fix the bug. In our study we focus on issues that are resolved (e.\,g., in Figure~\ref{fig:timeline}, $created(I_1) = 0$ and $resolved(I_3) = 10$). The function $committed: C \rightarrow \mathbb{N}, C \in \mathcal{C}$ returns the time stamp at which a commit was submitted to the version control system. e.\,g. $committed(C_6) = 8$ in the example.

\textbf{Temporal relations} Considering a non linked commit $C \in \mathcal{C}$, the temporal  structure imposes several constraints on the possible link candidates $I \in \mathcal{I}$. The following three cases exist.
\begin{enumerate}[leftmargin=0.5cm]
\item $committed(C) < created(I)$: Due to causality, the commit $C$ is not considered to be a link candidate for $I$ 
(e.g., in Figure~\ref{fig:timeline}, $C_2$ is not a link candidate for $I_2$).

\item $created(I) \leq committed(C) \leq resolved(I)$: This situation depicts the usual development work flow. After issue creation, the developers modify the source code and submit commits in order to resolve the issue. These commits are traced to the issue. Eventually, the issue is resolved, and in this example, no further commits are made to the issue (e.g., in Figure~\ref{fig:timeline}, the non linked commit $C_6$ is a link candidate for $I_3$).

\item $resolved(I) < committed(C)$: Intuitively, in this situation a trace link from $C$ to $I$ is not considered, since $I$ was already resolved before the commit occurred. However, this situation is not uncommon as Table~\ref{tab:late_commits} shows. The obvious reasons might be, that a developer forgot to submit the commit before resolving the issue. Another might simply be clock differences between the unconnected, decentralized systems used by Jira and Git which prevents strict time comparisons.
\begin{table}
 \renewcommand{\arraystretch}{0.8}
  \caption{Properties of linked commits. (1) The distribution of commits linked to issues after issues resolution along with the median time. (2) The average file overlap of consecutive commits linked to the same issue.}
  \label{tab:late_commits}
  \begin{tabular}{@{}l r@{\hskip 6pt}r@{\hskip 6pt}r@{\hskip 6pt}r@{\hskip 6pt}r@{\hskip 6pt}r@{}}
    \toprule
     &              \multicolumn{6}{c}{Project}       \\
    \cmidrule{2-7}
     & De   &   Dr &    Gr &    In &    Ma &    Pi \\
    \midrule
    \multicolumn{7}{l}{\hspace{-5pt}(1) Commits linked to already resolved issues}\\
    \ \ \ Number & 136  &  207 & 2,648 &  244 &  847 &  100 \\
    \ \ \ Median time after resolved    & 150h  &   60h &    5h &   19h &    5h &   60h \\
    \midrule
    (2) Avg. file commit overlap   & 0.35 & 0.35 & 0.71 & 0.33 & 0.40 & 0.45 \\
    \bottomrule
  \end{tabular}
  \vspace{-12pt}
\end{table}

In project \textsc{Derby}, there is sometimes a large discrepancy between the time at which an issue is resolved and the last commit that traces to it. For example, the improvement \texttt{DERBY-6516}\footnote{\url{https://issues.apache.org/jira/browse/DERBY-6516}} was resolved as fixed on 20/Mar/14; yet, on 4/Apr/14 a commit (\texttt{78227e4}\footnote{\url{http://goo.gl/j3WYd6}}) was submitted and linked to this improvement. However this scenario is quite rare, affecting only 136 commits. Interestingly, in both the \textsc{Groovy} and \textsc{Maven} projects, the median time difference for late commits is much lower (only 5 hours), but affects a huge number of commits. For example, in the \textsc{Maven} project, we observed that between 2005 and 2015 there was a constant offset between issue resolution and corresponding commit from either five or six hours as illustrated in \texttt{MNG-221}\footnote{\url{https://issues.apache.org/jira/browse/MNG-221}} (from 2005), \texttt{MNG-2376} (from 2008), and \texttt{MNG-5245} (from 2012).
\end{enumerate}
These temporal constraints limit the potential pairs of candidate links between non linked commits and non linked improvements and bugs.

\textbf{Structural relations} Table~\ref{tab:commit_to_issue_linkage} reveals (in row 1:n) that often multiple commits $C_a, C_b, \ldots$ are required in order to solve an issue $I$. Ideally all of these commits are traced to the respective issue. However, often only one commit in this series is explicitly linked to $I$. In \cite{DBLP:conf/iwpc/SchermannBPLG15} the other commits in this series are termed \emph{phantoms}. All commits in the series may share commonalities. In addition to their succession in time, the commits may modify a similar set of source code files since they are related to the same issue. We define a function $overlap(C_a, C_b) = \frac{mod(C_a) \cap mod(C_b)}{\max(|mod(C_a)|, |mod(C_b)|)}$ with $C_a, C_b \in \mathcal{C}$. For example, in Figure~\ref{fig:timeline} the overlap of $C1$ and $C2$ is $overlap(C_1, C_2) = \frac{1}{2}$, and $overlap(C_3, C4) = 0$. As shown in Table~\ref{tab:late_commits}, the average overlap of consecutive commits linked to the same issue varies among the projects. For example in \textsc{Derby}, the average overlap is 0.35 meaning, that, on average, one out of three files are the same for commits in a series. The highest number, 0.73, is achieved in \textsc{Groovy}, where a few files are changed  multiple times to implement an improvement or bug.  Three commits (\texttt{51d4fee}\footnote{\url{http://goo.gl/4YGjhe}}, \texttt{3d20737}\footnote{\url{http://goo.gl/AZcwmK}}, and \texttt{974c945}\footnote{\url{http://goo.gl/2TYhsz}}) were submitted between 4/Jan/2009 and 6/Jan/2009 all modifying one and the same source code file \texttt{DefaultGroovyMethods.java} and linked to improvement \texttt{GROOVY-3252}\footnote{\url{https://issues.apache.org/jira/browse/GROOVY-3252}}. This results in $overlap=1$ for each commit pair in the series.

Based on temporal closeness and overlap, there are indications that $C_2$ and $C_1$ may belong to a series of commits and thus could be traced to $I_1$. The situation may also occur forward in time, i.\,e. $C_7$ and $C_9$ may belong to a series because of temporal closeness and source file overlap and thus should be traced to bug $B_1$.

\subsection{Stakeholder Model}
Issues and commit artifacts both carry information about the author. The assignee of an issue also might be the person who contributes commits in solving the issue. In the studied scenario, there is no technical connection between the issue tracker Jira and the version control system Git. Thus we cannot rely on an available stakeholder model. Therefore we applied the following approach to identify individual developers in both systems. In each system, a developer is represented by a name and a login (nickname or email). In the first step, we separately collected all developers from the two systems and built two groups. In this step, we merged names if they used the same login and therefore were aliases for the same person. In the second step, we heuristically merged the two resulting developer lists and compared the names, in order to identify the same person in both systems. In order to fully protect user privacy and to comply with Github Privacy Requirements a unique number, \emph{user id}, was assigned to every developer.  The function $userid: u \rightarrow \mathbb{N}$ with $u \in \mathcal{C} \cup \mathcal{I}$ returns this user id for a given commit or issue.

\section{The Link Classifier}
\label{sec:LinkClassifier}
Our goal was to create a classifier that could identify issues associated with a commit.  The classifier was therefore trained to predict whether any issue-commit pair should be linked or not.

\subsection{Attributes of the Commit-Issue Relation}
Based on the artifact model introduced in the previous section, we identified 18 attributes per instance. These attributes fall into two categories, process-related information and textual similarity between artifacts using information retrieval techniques.\\

\noindent\textbf{Process-Related Attributes}\\[-6pt]

\noindent We consider the following 16 process-related factors to model the relationship between commits, source code files, and issues. These factors capture stakeholder-related, temporal, and structural characteristics of the candidate pair $(C, I)$ with $C \in \mathcal{C}, I \in \mathcal{I}$:

\textbf{Stakeholder-related information, $a_{1\ldots 3}$:} We capture the identities of the committer as $a_1 = userid(C)$ and the assignee of the issues as $a_2 = userid(I)$. Additionally, we marked as a binary attribute whether the two are identical as $a_3 = 1\, (\textrm{if}\,userid(C) = userid(I),\, 0$ otherwise). 

\textbf{Temporal relations between issue and commit, $a_{4\ldots 7}$:} Based on  temporal properties of issue and commit, we calculated $a_4 = committed(C) - created(I)$ and $a_5 = resolved(I) - committed(C)$. Additionally, we capture as $a_6$ whether $created(I) \leq committed(C)\qquad$ $\leq resolved(I)$, i.\,e. whether $C$ was committed during the active development time of $I$. Furthermore, we capture close commits in relation to issue resolution as $a_7 = |a_5| < \epsilon$. We set $\epsilon = 2.5$ days, derived from observing that late commits occur on average within 5 and 150 hours of the issue resolution for the studied projects (see Table~\ref{tab:late_commits}). For example in Figure~\ref{fig:timeline}, the pair $(C_6, I_3)$ yields $a_4 = 1$, $a_5 = 2$ and $a_6 = 1$.

\textbf{Closest previous linked commit, $a_{8\ldots 10}$:} We capture the set of previous commits linked to $I$ as $C_{prev} = \{C_x| is\_linked(C_x, I) \wedge committed(C_x) < committed(C)\}$. If non-empty, the commit $C_p \in C_{prev}$ with the largest commit time stamp is taken and used to calculate $a_8 = committed(C) - committed(C_p)$, $a_{9} = overlap(C_p, C)$, and $a_{10} = userid(C_p)$. For example in Figure~\ref{fig:timeline},  the pair $(C_2, I_1)$ yields $C_{prev} = \{C_1\}$ and thus $C_p = C_1$, $a_8 \approx 1$, and $a_9 = \frac{1}{2}$.
    
\textbf{Closest subsequent linked commit, $a_{11\ldots 13}$:} Analogous to the closest  previous linked commit, we capture subsequent commits $C_{next}$. We capture $C_{next} = \{C_x| is\_linked(C_x, I) \wedge committed(C) < committed(C_x)\}$ and selected $C_{n}$ with the minimal commit time to calculate $a_{11} = committed(C_n) - committed(C)$, $a_{12} = overlap(C_n, C)$, and $a_{13} = userid(C_n)$. For example in Figure~\ref{fig:timeline}, the pair $(C_7, B_1)$ yields $C_{next} = \{C_9\}$, $C_n = C_9$, $a_{11} \approx 2$, and $a_{12} = \frac{2}{3}$.

\textbf{Number of issues and existing links, $a_{14\ldots 16}$:} We calculate the set of existing issues at time $committed(C)$,\hfill\\ $I_{exist} = \{I_x | created(I_x) \leq committed(C) \leq resolved(I_x) \wedge I_x \in \mathcal{I} \}$ and capture its cardinality as $a_{14} = |I_{exist}|$. Taking $I_{exist}$, we derive $I_{user} = \{I_x| I_x \in I_{exist} \wedge userid(I_x) = userid(I)\}$ representing non-resolved issues for the assignee of $I$ at that instant in time and capture its size in $a_{15} = |I_{user}|$. With $a_{16} = |\{C_x | is\_linked(I, C_x) \wedge committed(C_x) < committed(C), \forall C_x \in \mathcal{C}\}|$ we capture the number of links to $I$ before commit $C$. For example in Figure~\ref{fig:timeline}, considering pair $(C_7, B_1)$, $I_{exist} = \{B_1, I_3, I_4\}$ and thus $a_{14} = 3$.\\

\noindent\textbf{Textual Similarity Attributes}\\[-6pt]

\noindent We leveraged information retrieval methods to compute textual and semantic associations between commit messages, source code files, and issues. We explored three primary techniques for computing textual similarity $sim$. These were the Vector Space Model (VSM), VSM with N-Gram enhancements (VSM-nGram), and Latent Semantic Indexing (LSI) \cite{Antoniol:Recovering, DBLP:conf/iwpc/EaddyAAG08, DeLucia:ArtefManag}. 

In the VSM model, each document, i.e., commit message, issue description, and source code file, is treated as an unstructured bag of terms. Following common information retrieval techniques, documents are pre-processed to remove stop words, to stem words to their morphological  roots, and to split camel-case and snake-case words (e.g., optionsParser vs. options\_parser) into their constituent parts. Each document $d$ is then represented as a vector $\vec{d} =$ $(w_{1,d},w_{2,d},...,w_{n,d})$, where $w_{i,d}$  represents the term weight associated with term $i$ for document $d$. Each term $t$ is assigned a weight using a standard weighting scheme known as $tf-idf$ \cite{DBLP:journals/tse/HayesDS06}.  The cosine similarity between a pair of vectors in then computed as follows in order to estimate the similarity between two documents $d1$ and and $d2$:
\begin{align} \label{eq:Sim}
sim(d1,d2)=\frac{(\sum_{i=1}^n w_{i,d1} w_{i,d2})}{\left(\sqrt{\sum_{i=1}^n w_{i,d1}} \cdot \sqrt{\sum_{i=1}^n w_{i,d2}} \right)}
\end{align}

The N-Gram enhancement to VSM utilizes n-gram models \cite{witten2016data, cavnar1995using}. N-gram is a contiguous sequence of $n$ words in a document. Each document is again represented as a vector, but in this case, the vector is comprised of both the word and the n-grams it contains. The documents are preprocessed in the same way as the basic VSM. Based on initial experimentation, we set $n$ from 2 to 4, to include 2-gram, 3-gram and 4-gram sequences in the vector representations. The similarity between vectors was again calculated using the cosine measure (equation~\eqref{eq:Sim}) with $tf-idf$ schema as described above.   


We conducted an initial comparative study of Latent Semantic Indexing (LSI) \cite{Antoniol:Recovering, DBLP:conf/iwpc/EaddyAAG08, DeLucia:ArtefManag}, VSM, and VSM-nGram.  Based on an initial comparison of the results we selected the VSM-nGram approach for computing textual similarity scores. This was because we observed that VSM-nGram outperformed VSM on our datasets, and ran much faster than LSI. In fact, the computation time of LSI on our datasets was prohibitively slow with runtimes of up to 40 hours in some cases, and so we rejected it as impractical. Furthermore, several previous studies have shown that VSM tends to either outperform LSI on software engineering datasets or perform in equivalent ways \cite{DBLP:journals/tse/HayesDS06, DBLP:books/daglib/p/LuciaMOP12, DBLP:conf/msr/0004RCRHV16}.  A detailed comparison of trace retrieval techniques within our classifier is outside the scope of this research. Therefore, based on our initial analysis, we chose VSM-nGram to compute the following similarity attributes:

\textbf{Textual similarity of a commit and an issue, $a_{17}$:} The similarity between the commit message and the textual content of the issue (for both improvements and bugs) is captured as $a_{17} = sim(C, I)$ with $C\in\mathcal{C}, I\in\mathcal{I}$. 

\textbf{Textual similarity of committed source files and an issue, $a_{18}$:} For each commit-issue pair, the textual similarity between the content of the most similar committed source code file and the textual content of the issue is captured as  $a_{18} = \max\{sim(F, I) | \forall F \in mod(C)\}$, $C\in\mathcal{C}, I\in\mathcal{I}$.

\subsection{Studied Attribute Sets}\label{sec:features_schemes}
We studied the impact of the presented attributes in four subsets.
\begin{itemize}[leftmargin=*]

\item \textbf{Process} -- This set solely contains the process-related attributes, i.\,e. $A_{Structure} = \{a_{1}\ldots a_{16}\}$. It studies the impact of all process-related attributes without considering textual similarity.

\item \textbf{Similarity} -- This set consists of the attributes\\ $A_{Sim} = \{a_{6}, a_{17},a_{18}\}$. It solely considers textual similarity between commit and issue given the constraint that the issue existed at the time of the commit.

\item \textbf{All} -- This set, $A_{all} = \{a_{1}\ldots a_{18}\}$, contains all process, similarity and stakeholder related attributes.

\item \textbf{Auto} -- This set, $A_{auto} \subseteq A_{all}$, addresses potential correlations and dependencies among attributes. It contains an automatically selected subset derived by considering the individual predictive ability of each attribute along with the degree of redundancy between them. We implemented the redundant attribute removal process based on Weka's inbuilt auto-selection feature \cite{Hall1998}.
\end{itemize}


\subsection{Dataset Profiles and Splits}
We aim to classify links between commits and improvements and between commits and bugs. We therefore construct two distinct profiles for each project, constructed from process and similarity attributes per commit-issue pair.
\begin{align*}
Profile_{p} &= (S_{p, train},\, S_{p, test}) \qquad p \in \{Bug, Imp\}
\end{align*}
Each profile consists of a distinct training and a testing set. We applied the following procedure per project to create instances of candidate commit--issue pairs for the training set $S_{p, train}$ as well as the testing set $S_{p, test}$.
\begin{align*}
S_{p, t} &= \{(C, I)\,|\, is\_candidate(C, I)\} && C \in \mathcal{C}_{p, t}, \, I \in \mathcal{I}_{p, t} \\
    & p \in \{Bug, Imp\} && t \in \{train, test\}
\end{align*}
with the function $is\_candidate$ defined as
\begin{align*}
is\_candidate(C, I) &= created(I) \leq committed(C) \\
       &\wedge committed(C) \leq resolved(I) + \epsilon
\end{align*}

The function limits the number of candidate commit/issue pairs according to causality. A link candidate is never considered between a commit if the issue has not been created at the time of the commit. Secondly, $\epsilon$ assures that a commit is not unboundedly considered as a candidate for issues resolved in the past. Based on an analysis of commits onto closed issues (see Table~\ref{tab:late_commits}), we found that the median commit time after the issue has marked resolved was between 5 and 150 hours for the studied projects and we decided to choose $\epsilon$ as 30 hours. The candidate sets $\mathcal{C}_{p, t}$ and $\mathcal{I}_{p, t}$ were then created as
\begin{align*}
    \mathcal{C}_{p, train} &= \{C\, |\, committed(C) \leq t_{split}\}  &&\qquad C\in\mathcal{C} \\
    \mathcal{I}_{p, train} &= \{I\, |\, resolved(I) \leq t_{split}\} &&\qquad I\in\mathcal{I}_{p} \\
    \mathcal{C}_{p, test} &= \{C\, |\, committed(C) > t_{split}\} &&\qquad p \in \{Bug, Imp\} \\
    \mathcal{I}_{p, test} &= \{I\, |\, created(I) > t_{split}\}. &&
\end{align*}

The parameter $t_{split}$ defines the point in time, which splits the training and test set. We choose a 80\% -- 20\% split and calculated $t_{split}$ as follows. First, we ordered the improvements in the respective project according to their creation date in ascending order. We selected the improvement $I_{split} \in \mathcal{I}_{Imp}$, which divides this sequence into 80\% and 20\% of all improvements where $t_{split} = resolved(I_{split})$, i.\,e. the resolution time of 80\% of all improvements.

Each commit-issue candidate in the profiles $Profile_{Bug}$ and $Profile_{Imp}$ forms an instance to train ($S_{p, train}$) or test ($S_{p, test}$) the classifier, where the test data (i.\,e. 20\% in every project) is distinct from the training data. For each instance, we calculate 18 attributes $a_{1\ldots 18}$ that characterize the relation between commit and issue. In addition, each instance is annotated with the known class (i.e. \emph{linked}, or \emph{non-linked}) as extracted from the projects' data. \emph{Linked} means that the developer had created an explicit tag from the commit to the issue, while \emph{non-linked} means that no such tag exists.


\subsection{Classifier Training}\label{sec:training_classifiers}
We investigated three different supervised learning classifiers for categorizing commit-issue pairs as \emph{linked} or \emph{non-linked}. These were N\"aive Bayes, J48 Decision Tree, and Random Forrest. We included N\"aive Bayes because even though the assumption of independence rarely holds in software project data, the algorithm has been demonstrated to be effective for solving similar prediction problems \cite{Kim:2011, d2012evaluating, Guo:2004, Falessi2017}. We utilized Weka's J48 decision tree with default pruning settings because of its previously reported effectiveness in other software engineering studies \cite{Guo:2004}. Finally, we included the Random Forest classifier because it has been shown to be highly accurate and robust against noise \cite{breiman2001random}, although it can be expensive to run on large data sets.

The profiles we created were severely unbalanced containing many more instances of \emph{non-links} than \emph{links}. Training against such unbalanced sets makes it likely that the classifier will favor placing instances into the majority class (i.\,e. in this case classifying all pairs as \emph{non-links}). We performed all experiments using Weka~\cite{weka:2009} and used the inbuilt sub-sampling feature to create balanced data sets. Given a fixed number of explicit \emph{links}, Weka randomly selects the same number of \emph{non-links}.
We trained each classifier in turn using the balanced sets $S_{Bug, train}$, and $S_{Imp, train}$ for each project and then evaluated the classifier against the respective unbalanced testing sets $S_{Bug, test}$ and $S_{Imp, test}$. To mitigate the random effects of sub-sampling, we repeated the training and testing 10 times and averaged the achieved results. We did not follow an ordinary 10-fold cross validation approach because several of the studied variables (e.\,g. attributes $a_{4} \ldots a_{7}$) reflect temporal sequences in the development, making it necessary to ensure that temporal sequencing between training and test data was preserved. For each technique the classifier returned a category (i.e. linked or non-linked) and also a score which we used to rank recommended links in order of likelihood.

\section{Data collection}
\label{sec:Data}
To prepare the data for training and testing the classifier, we performed a two step data collection process for each of the six projects.

\textbf{Step 1:  Analyzing project management and issue tracker system.} We implemented a collector to retrieve artifacts (i.\,e., improvements and bugs). All six projects use the Jira  project management tool offering a web-service interface. Our collector downloaded and parsed all artifacts. Using the artifact type, we filtered the artifacts to retrieve only bugs and improvements. Therefore we applied the following mapping from Jira types to our model: bug $\rightarrow$ bug, and improvement, enhancement $\rightarrow$ improvement. In both cases, the artifacts represented finished work, i.\,e. their status was "Resolved" or "Closed" and the resolution "Fixed" or "Done". 

\textbf{Step 2: Analyzing Source Control Management (SCM) system.} A second collector was implemented to download all source code changes and commit messages from each SCM repository (i.e., GIT).  We parsed the commit messages and applied the heuristic described in \cite{Bachmann:2009:SPD} to retrieve existing trace links from commits to bug reports and improvements based on searching and matching the issue keys in commit messages. Given the goals of our traceability experiment we excluded non-source code files related to documentation, and build automation based on their file name extensions. Additionally we analyzed file paths in order to exclude source code files implementing test cases. In the standard Maven directory layout\footnote{\url{https://goo.gl/D8uYaD}}, used by all six of our projects projects, source files are placed in sub-directories of \texttt{src/main} and tests as sub-directories of \texttt{src/test/java}.

The results of these two steps were stored in an archive per project, which is publicly available~\cite{ReproductionDataset2018}. Data were collected from each project until May 31\textsuperscript{th} 2017.

\section{Reconstructing Known Links}
\label{sec:KnownLinks}
We performed a two-phase evaluation. In the first phase we address RQ1 and RQ2 by exploring two different usage scenarios. The first uses the classifier as a \emph{recommender} system, to suggest a list of the most likely issues at the time a commit is submitted. Ideally this functionality would be integrated into the version control system and activated when the user presses the commit button. In this scenario, high recall is imperative, so that the relevant issue (if it exists) is included in the displayed list. The second experiment evaluates the case, in which the classifier is used to \emph{automatically augment} an existing set of trace links for a project. In this scenario, high precision is essential because links that are automatically added must exhibit high accuracy.  
In experiments related to both of these scenarios, we leveraged the existing links created by the project developers from explicit commit-issue tags as an ``answer set'' to train and evaluate our classifiers. Both experiments therefore evaluate whether the classifier would have been able to recommend or create a known link if the committer had forgotten to create its tag manually. We trained the three classifiers on the four attribute sets as described in Section~\ref{sec:features_schemes} and Section~\ref{sec:training_classifiers}.

Results were evaluated using commonly adopted traceability metrics. \emph{Recall} measures the fraction of relevant links that are retrieved while \emph{precision} measures the fraction of retrieved links that are relevant.  Finally, \emph{F-measure} measures the harmonic mean of recall and precision \cite{DBLP:conf/icse/ShinHC15, DBLP:journals/tse/HayesDS06,DBLP:conf/sigsoft/LoharAZC13,Antoniol:Recovering}.
We utilize two variants of the F-Measure -- namely $F_2$ which is weighted to favor recall, and $F_{0.5}$ which is weighted to favor precision.

\begin{table*}
\renewcommand{\arraystretch}{0.8}
\caption{Trace recommender evaluation using the different attribute sets and Random Forest (Scenario 1)}
\vspace{-2pt}
\label{tab:scenario_recommender}
\begin{tabular}{@{}ll rrr rrr rrr p{0.2cm} rrr@{}}
	\toprule
	&         & \multicolumn{3}{c}{Similarity} & \multicolumn{3}{c}{Process} & \multicolumn{3}{c}{Auto} & & \multicolumn{3}{c}{All} \\
	\cmidrule(l{2pt}r{2pt}){3-5}
\cmidrule(l{2pt}r{2pt}){6-8}
\cmidrule(l{2pt}r{2pt}){9-11}
\cmidrule(l{2pt}r{2pt}){13-15}
Project & Profile & P\textsuperscript{1}    &    R\textsuperscript{2} &       $F_{2}$\textsuperscript{3} & P    &    R &          $F_{2}$ & P    &    R &        $F_{2}$ & & P    &    R &         $F_{2}$ \\
\midrule
\multirow{2}{*}{Derby} & Bug              &         0.22 &         0.66 &            0.47 &             0.19 &             0.56 &                0.41 &      0.28 &      0.84 &         0.60 &  &   0.30 &     0.88 &        \textbf{0.63} \\
 & Improvement      &         0.23 &         0.69 &            0.49 &             0.29 &             0.87 &                0.62 &      0.31 &      0.92 &         0.66 &  &   0.32 &     0.95 &        \textbf{0.68} \\
\multirow{2}{*}{Drools} & Bug             &         0.27 &         0.80 &            0.58 &             0.33 &             0.97 &                0.70 &      0.34 &      1.00 & \textbf{0.72}&  &   0.34 &     1.00 &        \textbf{0.72} \\
 & Improvement     &         0.44 &         0.97 &            0.78 &             0.44 &             0.97 &                0.78 &      0.46 &      1.00 &         \textbf{0.81} &  &   0.46 &     1.00 &        \textbf{0.81} \\
\multirow{2}{*}{Groovy} & Bug             &         0.22 &         0.66 &            0.47 &             0.31 &             0.90 &                0.65 &      0.30 &      0.87 &         0.63 &  &   0.31 &     0.92 &        \textbf{0.66} \\
 & Improvement     &         0.30 &         0.87 &            0.63 &             0.29 &             0.83 &                0.60 &      0.30 &      0.88 &         0.64 &  &   0.33 &     0.95 &        \textbf{0.69} \\
\multirow{2}{*}{Infinispan} & Bug         &         0.27 &         0.81 &            0.58 &             0.30 &             0.91 &                0.65 &      0.30 &      0.91 &         0.65 &  &   0.31 &     0.92 &        \textbf{0.66} \\
 & Improvement &         0.31 &         0.92 &            0.66 &             0.33 &             0.96 &                0.69 &      0.33 &      0.98 & \textbf{0.71} & &    0.33 &     0.98 &        \textbf{0.71} \\
\multirow{2}{*}{Maven} & Bug              &         0.29 &         0.84 &            0.61 &             0.34 &             0.98 &                0.71 &      0.33 &      0.95 &         0.69 &  &   0.34 &     0.99 &        \textbf{0.72} \\
 & Improvement      &         0.31 &         0.89 &            0.65 &             0.31 &             0.89 &                0.65 &      0.34 &      0.97 & \textbf{0.70} & &    0.33 &     0.94 &        0.68 \\
\multirow{2}{*}{Pig} & Bug                &         0.30 &         0.90 &            0.64 &             0.32 &             0.97 &                0.69 &      0.33 &      0.98 &         0.70 &  &   0.33 &     0.99 &        \textbf{0.71} \\
 & Improvement        &         0.32 &         0.94 &            0.68 &             0.34 &             0.99 &                0.71 &      0.34 &      1.00 & \textbf{0.72} & &    0.34 &     1.00 &       \textbf{0.72} \\
\bottomrule
\end{tabular}\\
{\footnotesize Precision\textsuperscript{1}, Recall\textsuperscript{2}, $F_{2}$ Measure\textsuperscript{3}}
\vspace{-6pt}
\end{table*}

\textbf{Scenario 1: Recommending Issues to Assist Commits: }
The goal of this scenario is to create a short list with a maximum of three recommended links, to assist developers in tagging their new commits with issue IDs. Thus, we truncate the retrieved lists after the third rank and evaluate classifier performance in terms of precision, recall, and F2-measure at this point. F2-measure is selected because the objective of this scenario is to achieve high recall. The results for the best performing classifier Random Forest are shown in Table~\ref{tab:scenario_recommender}. The attribute set \emph{All} achieves an average recall of 96\% and average precision of 33\%, which in combination marks the best performance of the studied feature sets. An application of the Mann-Whitney U Test \cite{mann1947test} shows that the \emph{All} approach significantly ($p < 0.05$) outperforms the other attribute sets in terms of $F_2$ score. The other two classifiers also performed best when using the \emph{All} features sets. However, their achieved $F_2$ scores were significantly lower than that of Random Forest ($\overline{F_{2}} = 0.48$ for J48, and $\overline{F_{2}} = 0.6$ for N\"aive Bayes).
Generally, the values show that the Random Forest classifier is able to predict the one true link among the three recommended links. The attribute set \emph{Similarity} exhibits the lowest $F_{2}$ measure. The feature set \emph{Process} performs considerably well. This is notable, because it does not require resource intensive IR techniques to extract the necessary features $a_{17}$ and $a_{18}$. However, adding these features to the model results in overall better performance (see the \emph{All} attribute set). An exception within the results is the \textsc{Derby} project, which underperforms on all attribute sets. The low recall values indicate (e\,g., $0.56$ for \emph{Process}) that the correct link is not in the ranked list for one out of two commits. 

\begin{table*}
  \renewcommand{\arraystretch}{0.8}
  \caption{Fully automated trace link augmentation using the different attribute sets and random forest (Scenario 2)}
  \label{tab:scenario_automatic_augmentation}
  \begin{tabular}{@{}ll rrr rrr rrr p{0.2cm} rrr@{}}
    \toprule
    &         & \multicolumn{3}{c}{Similarity} & \multicolumn{3}{c}{Process} & \multicolumn{3}{c}{Auto} & & \multicolumn{3}{c}{All} \\
    \cmidrule(l{2pt}r{2pt}){3-5}
    \cmidrule(l{2pt}r{2pt}){6-8}
    \cmidrule(l{2pt}r{2pt}){9-11}
    \cmidrule(l{2pt}r{2pt}){13-15}
    Project & Profile & P\textsuperscript{1}    &    R\textsuperscript{2} &       $F_{0.5}$\textsuperscript{3} & P    &    R &          $F_{0.5}$ & P    &    R &        $F_{0.5}$ & & P    &    R &         $F_{0.5}$ \\ \midrule
    \multirow{2}{*}{Derby}  & Bug          &         0.67 &         0.43 &              0.60 &             0.91 &             0.08 &                  0.29 &      0.97 &      0.07 &           0.28 &   &  0.98 &     0.10 &          0.37 \\
    & Improvement           &         0.76 &         0.45 &              0.67 &             0.90 &             0.27 &                  0.62 &      0.96 &      0.35 &           0.71 &  &   0.98 &     0.28 &          0.66 \\
    \multirow{2}{*}{Drools} & Bug         &         0.89 &         0.36 &              0.69 &             0.93 &             0.69 &                  0.87 &      0.94 &      0.73 &           0.89 &  &   0.94 &     0.67 &          0.87 \\
    & Improvement           &         0.90 &         0.57 &              0.81 &             1.00 &             0.33 &                  0.71 &      0.97 &      0.03 &           0.15 &   &  1.00 &     0.23 &          0.60 \\
    \multirow{2}{*}{Groovy} & Bug             &         0.61 &         0.34 &              0.53 &             0.81 &             0.57 &                  0.75 &      0.97 &      0.05 &           0.22 &   &  0.89 &     0.41 &          0.72 \\
    & Improvement        &         0.70 &         0.56 &              0.67 &             0.88 &             0.44 &                  0.73 &      0.85 &      0.63 &           0.80 &   &  0.90 &     0.58 &          0.81 \\
    \multirow{2}{*}{Infinispan} & Bug         &         1.00 &         0.00 &              0.00 &             0.85 &             0.63 &                  0.79 &      0.89 &      0.58 &           0.80 &  &   0.93 &     0.48 &          0.78 \\
    & Improvement &         0.92 &         0.66 &              0.85 &             0.89 &             0.61 &                  0.82 &      0.93 &      0.69 &           0.87 &   &  0.97 &     0.69 &          0.89 \\
    \multirow{2}{*}{Maven} & Bug              &         0.88 &         0.46 &              0.74 &             0.97 &             0.38 &                  0.74 &      0.94 &      0.45 &           0.77 &   &  0.99 &     0.37 &          0.74 \\
    & Improvement      &         0.84 &         0.65 &              0.79 &             0.83 &             0.38 &                  0.67 &      0.94 &      0.63 &           0.86 &  &   0.95 &     0.52 &          0.82 \\
    \multirow{2}{*}{Pig} & Bug                &         0.98 &         0.55 &              0.84 &             0.89 &             0.75 &                  0.86 &      0.99 &      0.79 &           0.94 &   &  0.99 &     0.83 &          0.95 \\
    & Improvement        &         0.98 &         0.74 &              0.92 &             0.92 &             0.85 &                  0.90 &      0.97 &      0.92 &           0.96 &   &  1.00 &     0.90 &          0.98 \\
    \bottomrule
  \end{tabular}\\
  {\footnotesize Precision\textsuperscript{1}, Recall\textsuperscript{2}, $F_{0.5}$ Measure\textsuperscript{3}}
\end{table*}

\textbf{Scenario 2: Fully Automated Augmentation of Trace links between Commits and Issues:} The classifier performance for the second scenario is evaluated in terms of precision, recall, and $F_{0.5}$measure, because the objective of this scenario is to achieve high precision. Results are reported in Table~\ref{tab:scenario_automatic_augmentation} for the Random Forest classifier, which performed best. A fully automated environment requires high precision, thus we defined a project-independent cut off point based on $score > 0.95$, which achieves a precision above $\geq 90\%$ across all projects when using the \emph{All} attribute set. The other classifiers, J48 and N\"aive Bayes, were unable to achieve the required precision. For Random Forest, the recall drops to $50\%$ on average as a consequence of required precision and thus only one out of two known links would be re-created. 
In project \textsc{Derby}, the recall for \emph{All} is $10\%$, and similar values for the \emph{Process}, and the \emph{Auto} sets are achieved. However, the attributes set only containing textual similarity attributes performs best resulting in the highest $F_{0.5}$ measure, which favors precision over recall. As in the previous evaluation scenario, structural attributes do not perform well on this project, which is further discussed in the next section.


\section{Constructing Unknown Links}\label{sec:evaluating_unkown_links}
\label{sec:UnknownLinks}
The previous experiment was designed to reconstruct known links. However, the real value of our classifier is in recommending tags for commits with no existing links. While we have strong, albeit not perfect, confidence that the explicitly linked pairs of commits and issues are correctly labeled; however the non-links constitute a combination of true negative links (i.e. correctly labeled non-links) and false negative links (i.e. incorrectly labeled non-links).  Of these, the false negatives represent the \emph{missing links} that we now target. These missing links result from cases in which a developer failed to associate a commit with an issue or created an incomplete set of tags. Previous studies have reported the difficulty of correctly classifying entities not represented in the original training set \cite{DBLP:conf/msr/ParkKRB12} and we therefore need to evaluate the ability of the classifier to detect previously missing links.


\begin{table}
\renewcommand{\arraystretch}{0.8}
  \caption{Average number of classified links between issues and a non-linked commit}
  \label{tab:scenario_false_positives}
\begin{tabular}{@{}l@{\hskip 2pt}lrrr p{0.1cm} r@{}}
\toprule
Project & Profile &  Similarity &  Process &  Auto & &  All \\
\midrule
\multirow{2}{*}{Derby} & Bug              &          8.03 &             15.50 &       6.29 & &     4.91 \\
 & Improvement      &          5.90 &              6.75 &       5.85 & &     3.65 \\
\multirow{2}{*}{Drools} & Bug             &          0.90 &              0.49 &       0.42 & &     0.37 \\
 & Improvement     &          0.35 &              0.22 &       0.15 & &     0.21 \\
\multirow{2}{*}{Groovy} & Bug             &          2.05 &              5.41 &       1.52 & &     2.53 \\
 & Improvement     &          2.51 &              2.71 &       2.03 & &     1.77 \\
\multirow{2}{*}{Infinispan} & Bug         &          0.00 &              2.63 &       1.74 & &     0.69 \\
  & Improvement &          1.85 &              1.08 &       0.88 & &     0.82 \\
\multirow{2}{*}{Maven} & Bug              &          0.83 &              0.54 &       0.73 & &     0.46 \\
 & Improvement      &          0.94 &              0.72 &       0.65 & &     0.61 \\
\multirow{2}{*}{Pig} & Bug                &          1.40 &              1.93 &       0.93 & &     1.20 \\
 & Improvement        &          0.89 &              1.56 &       1.00 & &     1.22 \\
\bottomrule
\end{tabular}
\end{table}
Since no answer set for the non-linked commits is available, we needed to perform a manual inspection of the proposed links. As a sanity check we first evaluated whether it would be plausible to classify links on these unknown parts. In all six projects commits with links are typically related to only one issue or to a very small number of them (see Table~\ref{tab:commit_to_issue_linkage}). Therefore, we count the average number of issues classified as links for each of the commits without any explicit link (see Table~\ref{tab:scenario_false_positives}). 
The classifier trained using attribute set \emph{All} identifies an average of $1.54$ issues per commit as a candidate link. Table~\ref{tab:issue_to_commit_linkage} and \ref{tab:commit_to_issue_linkage} characterize the current linking situation in the studied projects. Based on these values, we expect a value of $\approx 1.2$ links per commit. For example in project \textsc{Infinispan}, there were $2,223$ commits linked to bugs and $1714 + 254 = 1968$ bugs linked to commits, and thus $\approx 1.12$ commits per bug. The ratio for improvements is $1.32$. However, the classifier proposes $0.69$ bugs per commit and $0.82$ improvements for every non-linked commit. That means that our approach is conservative.
For project \textsc{Derby}, our approach underperformed. The existing ratio for linked commits per bug is $1675 / (273 + 1093) = 1.21$, as for the other projects. But the classifier suggests $4.91$. This may stem from the imbalance of non linked commits and non linked issues in the project. There are $2,202$ non linked issues and only $553$ non linked commits. However, the same imbalance of non linked issues and commits also exists in project \textsc{Pig}, but in this context, the classifier is unaffected.

This analysis shows that except for \textsc{Derby}, the classified number of links is plausible. However, it is not clear whether these links are correctly classified. To accomplish this goal we manually evaluated the correctness of a random selection of new links proposed by the classifier using the following systematic process for each project.  Steps 1-4 are independently performed by one researcher (data preparer), while steps 6-7 are performed collaboratively by four additional researchers (referred to as evaluators). No communication was allowed between the researcher creating the dataset and the four evaluators during this process.\\

\noindent\textbf{Data Set Construction for Missing Links}
\begin{enumerate}[leftmargin=0.5cm]
    \item[D1] Twenty commits without any explicitly tagged issues from the original data set for a given project were randomly selected and randomly divided into two groups $A$ and $B$. 70\% were placed into group A and 30\% into group B.  
    \item[D2] For commits in group $A$ the most highly ranked issue ID was selected as the candidate link, while for commits in group $B$ an issue tag that was not recommended by the classifier was selected. Group $B$ was added to mitigate evaluation bias and to ensure a mix of links and non-links in the evaluation set.
    \item[D3] A randomly ordered list of each commit-issue pair selected in the previous step was generated:
\end{enumerate}

\noindent\textbf{Human Evaluation of Proposed Links}
\begin{enumerate}[leftmargin=0.5cm]
    \item[H1] Four human evaluators worked together to classify the first five commit-issue pairs from one randomly selected project. They performed this task without any knowledge of whether the link was recommended by the classifier or not. The evaluators then worked individually to classify the next five commit-issue pairs in the list.
    \item[H2] The Fleiss kappa inter-rater agreement was computed. Fleiss's kappa assesses the likelihood of more than two raters agreeing when classifying items into a set number of categories \cite{Fleiss1971}. A kappa value of 1 means that all raters are in agreement, though a value above 0.4 indicates strong agreement. Evaluators discussed results for 20 commit-issue pairs with the aim of achieving consensus in classifying the pair as having a link or not. The Fleiss kappa value for this evaluation was approximately 0.5617, demonstrating the reliability of the evaluators to agree on the link status between a commit and issue. 
    \item[H3] As satisfactory inter-rater agreement was achieved, the remaining pairs of commit-issues were split amongst the evaluators and all pairs were evaluated.  The decisions made by the evaluators constitutes the ``answer set'' of previously unknown links against which the classifier is evaluated.
\end{enumerate}

 Due to the labor intensive nature of this analysis, we evaluated only three projects: \textsc{Derby}, \textsc{Drools}, and \textsc{Maven}. Recall and precision were computed by comparing the results returned by the classifier against the manually created ``answer set''. Results obtained for forty commit-issue pairs for each project are summarized in Table~\ref{tab:human_evaluation_results}.

\begin{table}
  \renewcommand{\arraystretch}{0.8}
  \caption{Links Recommended to Commits with no Tags}
  \label{tab:human_evaluation_results}
  \centering
  \begin{tabular}{@{}l@{\hskip 2pt}l@{\hskip 2pt}rrrrrrrr@{}}
    \toprule
    Project
    &Profile
    &L\textsuperscript{1}
    &NL\textsuperscript{2}
    &TP\textsuperscript{3}
    &FP\textsuperscript{4}
    &TN\textsuperscript{5}
    &FN\textsuperscript{6}
    &P\textsuperscript{7}
    &R\textsuperscript{8} \\
    \midrule
    \multirow{2}{*}{Derby}  &Bug&7&33&7&21&12&0&0.25&1.0\\
    &Imp&3&37&3&25&12&0&0.11&1.0\\
    \multirow{2}{*}{Drools} &Bug&7&33&7&21&12&0&0.25&1.0\\
    &Imp&8&32&8&20&12&0&0.28&1.0\\
    \multirow{2}{*}{Maven}  &Bug&3&37&3&25&12&0&0.11&1.0\\
    &Imp &2&38&1&27&11&1&0.04&0.5\\
    \bottomrule
  \end{tabular}\\
{\footnotesize Links\textsuperscript{1}, Non-Links\textsuperscript{2}, True Positives\textsuperscript{3},False Positives\textsuperscript{4}, True Negatives\textsuperscript{5}, False Negatives\textsuperscript{6}, Precision\textsuperscript{7}, Recall\textsuperscript{8}}
\end{table}

Results indicate that all projects except one returned a recall of 100\% (i.e. 100\%). The exception was \textsc{Maven} where recall of 100\% was achieved for commit to bug links, but only 50\% for commit to improvement links.  In this case, there were only two true links, and one of them was missed. This means that the classifier found the pair to be unconnected while the evaluator determined that a link did exist.  
The precision returned for each of the three projects for both bugs and improvement was lower than the precision returned in the earlier experiments with explicitly defined links. For example, in earlier experiments \textsc{Derby}'s precision was 0.30 for bugs and 0.32 for improvements.  However, these scores dropped to 0.25 and 0.11 respectively when the classifier was used to generate links for commits with no previously known issue tags.  Similar trends were observed for \textsc{Drools}.  However, precision dropped considerably for \textsc{Maven} returning 0.11 for bugs, but only 0.04 for improvements.
A potential explanation for the poor precision result in the \textsc{Maven} project is the fact that a majority of the commits represent code refactoring and in many cases were not associated with any issues at all - resulting in several false positive links.  This was also the case in other projects where several commits were not directly associated with any particular issue but addressed a more trivial task such as correcting a typo or adding a comment in java docs. These types of commit negatively impact overall precision.
\section{Related Work}
\label{sec:Related}
The most closely related work falls under the two areas of feature location and tracing bug reports to code.  

Feature location attempts to identify sections of source code related to a specific requirement or issue. Several authors have looked at static approaches based on information retrieval techniques. For example, Antoniol et al. \cite{Antoniol:Recovering} used a probabilistic approach to retrieve trace links between code and documentation. Hayes et al. used the Vector Space Model (VSM) algorithm in conjunction with a thesaurus to establish trace links \cite{DBLP:journals/tse/HayesDS06}.  Other studies applied Latent Semantic Indexing \cite{DeLucia:ArtefManag,DBLP:conf/icse/RempelMK13}, Latent Dirichlet Allocation (LDA) \cite{DBLP:conf/re/DekhtyarHSHD07, DBLP:conf/icse/AsuncionAT10}, or recurrent neural networks  \cite{DBLP:conf/icse/0004CC17} to integrate semantics or context in which various terms are used. Other researchers have combined results of individual algorithms \cite{lohar2013improving, DBLP:conf/re/DekhtyarHSHD07, GethersICSM}, applied AI swarm techniques \cite{DBLP:journals/re/SultanovHK11} and combined heuristic rules with trace retrieval techniques \cite{DBLP:journals/jss/SpanoudakisZPK04, 6636704, DBLP:conf/re/Cleland-HuangMMA12}. Our approach leverages information retrieval to compute similarity between various types of issues, commit messages, and code. We investigated the use of LSI but rejected it for a pragmatic reason that it had a long execution time, and further, that prior studies have not shown it to outperform VSM.  Ultimately we adopted a VSM-based approach, that outperformed basic VSM and integrated natural language concepts.

Researchers have also integrated structural analysis of the code to support feature location \cite{DBLP:conf/kbse/RasoolM11,DBLP:conf/icse/McMillanHPCM12,DBLP:conf/csmr/PanichellaMMPOPL13}. We did not include this in our current classifier; however, we will consider it in future work. Structural analysis may be especially helpful for finding additional classes that are related to an issue or bug. In less closely related work, researchers investigated the use of dynamic analysis for feature location \cite{DBLP:conf/icsm/KuangMHGHJE12,DBLP:journals/smr/KuangMHGHLE15,DBLP:conf/wcre/KuangNHRLEM17}. Furthermore, Eisenbarth et al., \cite{eisenbarth2003locating} presented a technique combining dynamic and static analyses to rapidly focus on the system's parts that relate to a specific set of features. 

Our work focuses not only on feature requests (i.e. improvements), but also tracing bugs to code. Canfora et al. used information retrieval techniques to identify files that were created or changed, in response to a Bugzilla ticket \cite{canfora2005impact}. They identified files changed in response to similar bug reports in the past, using standard information retrieval techniques.  Kim et al. predicted which source code files would change as a result of bug-fix requests \cite{kim2013should} using Mozilla FireFox and Core code repositories as their corpus in tandem with the public Bugzilla database. They first trained a classifier to recognize `usable' versus `non-usable' bug reports, and then using the bugs classified as usable, trained a second classifier to identify impacted classes.  Our approach differs from their work in that our goal is to generate links directly from commits to issues so that we can make direct recommendations to users if they forget to tag a commit. Our goal is therefore to create trace links as the commits are made so that developers can accept or reject them in order to create a set of trusted links. In \cite{DBLP:conf/iwpc/SchermannBPLG15}, the authors  proposed two heuristics, Loners and Phantoms, to infer trace links between commits and issues. We incorporate their concepts as one attribute in our classifier. 

\section{Threats to Validity}
\label{sec:Threats}
There are several potential threats to the validity of our study.
\par\noindent\textbf{Internal Validity} We split the available data set for each project into 80--20\% of the issues retaining the temporal ordering of the project. Choosing another split point may produce different evaluation results. We considered explicitly only ``resolved'' bugs and improvements, assuming that all required source code modifications had already taken place. It may be possible that the process of resolving an issue does not manifest in commits. We tried to mitigate this, by focusing on commits marked as ``Fixed'' or ``Resolved''; however, some commits might intentionally not address an issue due to their triviality. This was evidenced in our final experiment, where our classifier recommended links  even though no links existed. Furthermore, our study focused on improvements and bugs, as these were the predominant types of instances in our projects; however, we observed comparable commit link patterns for other issues types, suggesting that our approach would generalize.

\par\noindent\textbf{External Validity} Our study focused solely on open-source projects. A potential threat to external validity arises when we want to generalize our findings to a wider set of project, including commercial development. We have observed evidence of similar tagging practices in our own industrial collaborations, and therefore expect similar results. However internal company regularities might influence commit practices, and thus the overall applicability of our approach is an open question. Another threat that might limit the generalizability of our results is the use of only one combination of issue tracking system (Jira) and version control system (Git). Other tools and platforms might encourage and/or provide different linking behavior. 

\section{Conclusion}
\label{sec:Conclusion}
In this paper, we studied the interlinking of commits and issues in open source development projects. An analysis of six large projects showed that on average only 60\% of the commits are linked to issues. This incomplete linkage fundamentally limits the establishment of project-wide traceability. To overcome this problem, we propose an approach that trains a classifier to recommend links at the time commits are made and also augments an existing set of commits and issues with automatically identified links. We identified structural, temporal, stakeholder-related and textual similarity factors as relevant information for automating this task and derived 18 attributes to quantify the relation between commit--issue pairs. A Random Forest classifier performed best on the trained attributes. We evaluated this trained model through conducting four different experiments. Two experiments studied classification performance for recommending links upon a new commit as well as for automatically augmenting missing links. We found that the classifier yielded on average $96\%$ recall in a short list of three recommendations and could on average automatically augment every second link correctly with an average error of $4\%$. Finally, we manually constructed a small answer set of links from the set of previously unlinked commits and showed that the classifier returned high recall results averaging 91.6\% and precision of 17.3\%.

\section*{Acknowledgments}
The work was partially funded by the German Ministry of Education and Research (BMBF) grants: 01IS14026A, 01IS16003B, by DFG grant: MA 5030/3-1, and by the EU EFRE/Th{\"u}ringer Aufbaubank (TAB) grant: 2015FE9033. It was also funded by the US National Science Foundation Grant CCF:1319680.

\bibliographystyle{ACM-Reference-Format}
\bibliography{trace} 

\end{document}